# Simulating Anisotropic Thermal Conduction in Supernova Remnants II : Implications for the Interstellar Medium

By

Dinshaw S. Balsara (dbalsara@nd.edu), Anthony J. Bendinelli (abendine@nd.edu), David A. Tilley (dtilley@nd.edu), Andrew R. Massari (amassari@nd.edu), J. Christopher Howk (jhowk@nd.edu)

Physics Department, University of Notre Dame

**Mailing Address:**

Physics Department

College of Science

University of Notre Dame

225 Nieuwland Science Hall

Notre Dame, IN 46556

**Phone :** (574) 631-9639

**Fax :** (574) 631-5952

**Abstract**

We present a large number of 2.5d simulations of supernova remnants expanding into interstellar media having a range of densities, temperatures and magnetic field strengths. The simulations include equilibrium cooling and anisotropic, flux limited thermal conduction along magnetic field lines. The volume of hot gas produced during the remnant's evolution is shown to be strongly influenced by the inclusion of thermal conduction, supporting prior results by Slavin & Cox (1993) and Tilley & Balsara (2006). The magnetic field has also been shown to play an extremely important role in reheating the gas at later epochs when the hot gas bubble collapses on itself. Low density, strongly magnetized runs show the greatest effect of this reheating.

The four-volumes and three-areas of gas with characteristic temperatures that cause it to emit in OVI, OVII and OVIII have also been catalogued and their dependence on interstellar parameters has been documented. The results reveal the importance of magnetic tension forces as well as the anisotropic thermal conduction along field lines for the production of these ions. Simulated luminosities and line widths of OVI, OVII and OVIII as well as their dosages have been catalogued.

Simulated line widths of radioactive species, $^{26}$Al and $^{60}$Fe, ejected by supernovae have also been catalogued and found to be less than 200 Km/sec for most of the remnants' evolution. These results enable us to understand why INTEGRAL has thus far been unable to detect very large line widths for these radioactive species in certain star forming regions.

**I) Introduction**

Chandra, XMM and FUSE have recently made it possible to study the hot gas content in our interstellar medium (ISM) in considerable detail, see Yao & Wang (2005, 2007), Bowen et al (2006), Savage & Lehner (2006), Dixon, Sankrit & Otte (2006). The soft x-rays observed in our Galaxy, McCammon et al (1983), Snowden et al (1997), are emitted by the hot gas from individual supernovae or from OB associations in which winds and supernova shells collide to produce gas that is even hotter than in isolated supernova remnants (SNRs). This hot gas also escapes the disk to form the hot gas halo of our Galaxy (Yao & Wang 2005, Wakker et al. 2003, Savage et al. 2003). Snowden & Petre (1994) have also find evidence for soft x-rays from hot gas in the LMC. While some of this x-ray emission comes from star-forming complexes, the rest comes from relatively quiescent regions, a conclusion also drawn from studies of LMC O VI (Howk et al. 2002; Lehner & Howk 2007), indicating an origin in a prior generation of SNRs that are not detectable any more.

Supernova remnants dominate the energy injection in the turbulent ISM. Winds from OB associations are a sub-dominant source of energy and provide as much as 10% of the mechanical energy that is supplied by SNRs. Thus the evolution of SNRs has a major influence on several observable features of our ISM. As shown in the one-dimensional studies of Slavin & Cox (1992,1993, SC92 and SC93 henceforth), thermal conduction (TC, henceforth) greatly influences the energy transport within the SNRs and also within the ISM at large. It is, however, not possible to include all the effects of dynamically active magnetic fields in one-dimensional simulations of SNRs, nor is it possible to include the anisotropic effects that such fields introduce in the evolution of SNRs. It has recently become possible to include anisotropic TC in magnetized simulations of SNRs, see Tilley & Balsara (2006, TB henceforth), Tilley, Balsara & Howk (2006, TBH henceforth) and Balsara, Tilley & Howk (2007, BTH henceforth). The purpose of the present study is, therefore, to simulate the evolution of SNRs in magnetized ISMs with TC. The goal of this paper is to extract observable diagnostics that are relevant to the ISM. The present paragraph has set the stage for the rest of this introduction by showing the importance of TC in determining the evolution of individual SNRs and their contribution to the evolution of the ISM at-large. In subsequent paragraphs we discuss the supernova-driven turbulent ISM, highlighting the role of TC in those systems.

Studies of SNRs without TC had shown that the interiors of remnants form an extremely hot under-dense bubble, see Cioffi, McKee & Bertschinger (1988, CMB henceforth). Bubbles with extremely high temperatures and very low densities radiate very inefficiently. As a result, models for such bubbles have not been able to explain the soft x-ray emission observed by McCammon et al (1983). Nor can they explain the emissivity in high-stage ions such as O VI, N V, C IV and Si IV (Park et al. 2007, Kim et al. 2007, Shinn et al. 2006, Blair et al. 2002). Explaining such emission requires gas in the range spanning $3 \times 10^5$ K to a few times $10^6$ K, with the O VI data favoring the lower part of that temperature range and the soft x-ray data favoring the higher part of that range. (FUSE has supplied more recent measurements of the hot gas and high-stage ions, see Oegerle et al 2005, Bowen et al 2005 and Savage & Lehner 2006.) In view of the previously-presented data, SC92 proposed that TC could transport heat away from the remnants' interior lowering the temperature in the interior of the remnants to the range where they could become efficient emitters of soft x-rays and the UV lines of high-stage ions. Unlike radiative cooling which can reduce the thermal energy of radiating gas, TC only transfers thermal energy from hotter parcels of gas to colder ones. Since the interiors of remnants evolve isobarically, the reduced temperature also results in an increased density, further increasing the emissivity from the hot bubble. A natural corollary of such a scenario is that the hot bubble cools more efficiently in the later stages of its evolution because it is denser. Thus SNRs with TC should fade out faster, resulting in smaller volume filling factors for the hot gas observed in the ISM. SC93 presented several one-dimensional models of SNRs with flux-limited TC, non-equilibrium ionization and magnetic fields. In SC92 and SC93 the magnetic pressure was treated as a scalar variable that varies as the square of the density and the magnetic tension forces were ignored. The magnetic field also makes the TC anisotropic and, owing to their one-dimensional approximation, SC93 ignored the effects of anisotropic TC. Because our present calculations are multidimensional, we are able treat the magnetic tension forces correctly, and we include the role of magnetic fields in making the thermal conduction anisotropic. In this paper we only include the effects of equilibrium ionization and cooling, the effects of non-equilibrium ionization will be examined in a subsequent paper. The data that we generate for the high-stage ions in this paper will, therefore, serve as a foil for comparison with subsequent work that includes non-equilibrium ionization.

Field, Goldsmith & Habing (1969) demonstrated that the form of the interstellar cooling and heating curves allowed a range of pressures for which two isobaric phases could exist in thermal equilibrium. Cox & Smith (1974) emphasized the role of SNe in producing the hot interstellar medium. Drawing on the ideas of these earlier works, McKee & Ostriker (1977; henceforth MO) presented a three phase model for the interstellar medium. More recently, Wolfire et al (1995a,b, 2003) emphasized the importance of polycyclic aromatic hydrocarbons or small grains to the thermal balance of the warm and cold phases. The relevance of SN-driving in producing interstellar turbulence and the role of that turbulence in determining the structure of the ISM has recently been reviewed in Scalo & Elmegreen (2004) and Elmegreen & Scalo (2004). The MO model has recently been subjected to theoretical analysis, Ferriere (1998) and Oey (2003) and to tests via direct numerical simulations, Korpi et al (1999), Balsara et al (2004), Avillez & Breitschwerdt (2004), Mac Low et al (2005) and Balsara & Kim (2005). Observational tests of such a model for the multi-phase ISM include estimates of the hot gas filling fractions, Brinks & Bajaja (1986) and Wang et al (2001), and the measurement of high-stage ions, Bowen et al (2005) and Savage & Lehner (2005). In this paper we catalogue the hot gas filling fractions as well as the dosages of high-stage ions produced by simulated SNRs that propagate through ISM models with various densities and temperatures.

The energy transport in such an ISM plays an important role in the distribution of gas phases in our Galaxy. Balbus & McKee (1982), Balbus (1986), Borkowski, Balbus, & Fristrom (1990) and Begelman & McKee (1990) have pointed out the importance of TC in transporting energy across the phases in the ISM. The coefficient of thermal conduction has a strong temperature dependence. As a result, TC is more likely to influence energy transport between the hot and warm phases (TB) than between the warm and cold phases (Audit & Hennebelle 2007). Of late, several authors have claimed that the turbulent transport of energy in the ISM may be more important than the transport of energy via TC, Cho et al (2003), Balsara & Kim (2005), Avillez & Breitschwerdt (2005). Implicit in these analyses is the viewpoint that hot gas bubbles from very old SNRs will be advected by the interstellar turbulence. The advective transport of the hot gas, and the energy that gets transported in the process, is then deemed to be faster than the rate at which TC transports energy. We point out, however, that there is another dimension to the argument. It stems from the fact that TC might also change the temperature distribution in the early and intermediate stages of SNR evolution. If it does then it will also influence the temperature structure within younger SNRs. As a result, it changes the distribution of x-ray emitting gas as well as the gas that forms the high-stage ions. Thus extracting diagnostics that rely sensitively on the temperature structure of the gas might be risky if TC is excluded. Tilley & Balsara (2006, TB henceforth) evaluated the filling factors for hot gas at several physically representative temperatures and found that the inclusion of TC dramatically changes the amount of hot gas that is produced. In this paper, we further examine how TC influences the hot gas filling fraction during all stages in the evolution of a SNR.

The INTEGRAL satellite has detected the line produced by the β-decay of radioactive $^{26}$Al and also placed upper bounds on its line width. While the β-decay line of $^{60}$Fe has also been detected by INTEGRAL, a measurement of the line width might only occur in a subsequent mission. Both these radioactive isotopes are made in core collapse supernovae and $^{26}$Al has an additional source in Wolf-Rayet stars. The line widths of $^{26}$Al have been measured towards various sources and also in the Galactic disk (Diehl et al 2006). They report an upper limit on line widths of 340Km/sec (Diehl 2006) and attribute the high line widths to interstellar turbulence. We note that the half life of $^{26}$Al is ~ 1 Myr , making it difficult for the $^{26}$Al to be uniformly mixed in with the ISM. If it is indeed mixed in with the warm phase of the ISM, it would also have a turbulent velocity of ~ 10 Km/sec. In this paper we, therefore, explore the possibility that expanding SNR bubbles, especially in their late stages of evolution, might provide the large line widths.

In Section II we describe the models that were run. In Section III we present the ISM-related observables that can be extracted from the simulations. Section IV provides a discussion. Section V presents conclusions.

**II) Description of Models**

The simulation code is described in Balsara (1998a,b), Balsara & Spicer (1999a,b) and Balsara (2004). The cooling function was drawn from MacDonald & Bailey (1981). The anisotropic, flux-limited thermal conduction draws on the work of Cowie & McKee (1977) and Balbus & McKee (1982) with a further description of the numerical scheme in BTH. The SNR simulations were carried out on a cylindrical mesh with the toroidal direction suppressed and are described in detail in BTH. We choose various values for $R_{outer}$, the outer boundary of the computational domain in the r and z directions and catalogue it in Table 1. The number of zones in the simulations, $N_{zones}$, is also catalogued in Table 1. In each case, the simulations were run until the hot gas bubble collapsed. The final time $t_{final}$, out to which the simulations were run is also catalogued in Table 1.

The ambient ISM was always chosen to be quiescent with a constant density $\rho_{ism}$ (measured in units of $10^{-24}$ gm/cm$^3$), a constant temperature $T_{ism}$ (measured in Kelvins) and a constant magnetic field $B_{ism}$ (measured in µGauss). To simplify later analyses, we also catalogue $v_{ism}$ (measured in Km/sec), the fast magnetosonic speed in the ISM. In each case, $10^{51}$ ergs (FOE), or a small multiple of it was injected using the prescription of BTH. Some of the runs were repeated without TC to enable us to assess the role that TC plays in the simulations. Table 1 shows the parameters used for the runs.

While half of the runs were presented in BTH, for the sake of completeness we catalogue them all here. The nomenclature of the runs is as follows: VL0e1, VL2e1, VL6e1 refer to very low density runs with $\rho_{ism} = 0.2 \times 10^{-24}$ gm/cm$^3$, magnetic fields of 0, 2 and 6 µGauss respectively and $10^{51}$ ergs of energy in the initial SNR. L0e1, L3e1, L6e1 refer to low density runs with $\rho_{ism} = 0.7 \times 10^{-24}$ gm/cm$^3$, magnetic fields of 0, 3 and 6 μGauss respectively and $10^{51}$ ergs of energy in the initial SNR. Runs L0e1, L3e1, L6e1 are done with ISM parameters that are currently considered standard. H0e1, H6e1, H20e1 refer to high density runs with $\rho_{ism}$ = 5.0X$10^{-24}$ gm/cm$^3$ , magnetic fields of 0, 6 and 20 μGauss respectively and $10^{51}$ ergs of energy in the initial SNR. VH0e1, VH6e1, VH20e1 refer to very high density runs with $\rho_{ism}$ = 20.0X$10^{-24}$ gm/cm$^3$ , magnetic fields of 0, 6 and 20 μGauss respectively and $10^{51}$ ergs of energy in the initial SNR. The similarly named runs that end in "e2" have 2X$10^{51}$ ergs of energy in the initial SNR.

**III) Extracting Observable Quantities From the Simulations**

In this Section we extract observationally relevant quantities from our simulations. Sub-section III.1 focuses on the evolution of the outer shock and the hot gas bubble. In Sub-section III.2 we obtain soft and hard x-ray fluxes. In Sub-section III.3 we extract observables, such as four-volumes, that are directly relevant to the measurement of the hot gas filling fraction in our ISM. In Sub-section III.4 we obtain diagnostics that are relevant to the measurement of high-stage ions in our ISM. In Sub-section III.5 we catalogue line widths of radioactive species, $^{26}$Al and $^{60}$Fe, that have recently been measured by the INTEGRAL satellite.

**III.1) Evolution of the Outer Shock and the Hot Gas Bubble**

The evolution of the hot gas bubbles in our simulations is governed by several physical processes that include radiative cooling, the distribution of thermal and magnetic energies and the role of thermal conduction in magnetized and un-magnetized environments. In order to develop a good understanding of the evolution of the hot gas bubble, we compare the distribution of thermal and magnetic energies in two of our simulations, L0e1 and L3e1, which differ only by the presence of the magnetic field. Fig. 1a shows the time evolution of the volume-weighted mean temperature of the hot gas bubble. We identified the hot gas bubble by restricting attention to gas with temperature in excess of 2X$10^5$ K. The solid line corresponds to L0e1 and the dashed line corresponds to L3e1. Fig. 1b shows the time evolution of the thermal and magnetic energy in the hot gas bubbles as well as the same in the shocked ISM. The solid line in Fig. 1b shows the thermal and magnetic energy in the hot gas bubble of L0e1, the dashed line displays the same for L3e1. By focusing on the shocked ISM gas that is not part of the hot-gas bubble, we can also identify the thermal and magnetic energy in the shocked interstellar gas. The dotted line in Fig. 1b shows that energy for L0e1 while the dot-dashed line shows the same for L3e1. Fig. 1a shows us that the average temperature in L0e1 is lower than that in L3e1 by a factor of up to two. The presence of a magnetic field in run L3e1 constrains the thermal conduction to be one-dimensional, i.e. along field lines. Consequently, we attribute the lower temperatures in L0e1 to the fact that thermal conduction operates multi-dimensionally in that run. Fig. 1b shows us that the thermal and magnetic energy in the hot gas bubble declines with time, as expected. Since the mean temperature of the hot gas bubble is higher in L3e1, we see that the thermal and magnetic energy in the hot gas bubble of L3e1 is also systematically higher that that of L0e1 for most of their evolution. As the shock propagates outwards, sweeping up progressively larger amounts of interstellar matter and magnetic field, we expect the energy of the shocked gas to increase with time. This is shown by focusing on the dotted and dot-dashed lines in Fig. 1b. The thermal energy in the shocked region is easily radiated away, especially at later times while the energy in the compressed magnetic field is retained locally. Consequently, we see that the thermal and magnetic energy in the magnetized L3e1 run is about a half order of magnitude greater than the corresponding energy in the non-magnetized L0e1 run. This trend becomes especially prominent at later epochs, where radiative losses dominate. Fig. 1b also shows us that the thermal and magnetic energy in the shocked interstellar gas exceeds that in the hot gas bubble at a few times $10^5$ years. By about a Myr we see that the thermal and magnetic energy in the shocked interstellar gas is greater than that in the hot gas bubble by more than an order of magnitude. At that point, we would expect the total pressure in the ambient medium to drive shocks into the hot gas bubble. Movies of the simulations bear out this expectation. The next few paragraphs will show that these times in the simulations also correspond to the times when the bubbles begin to collapse on themselves.

Fig. 2a shows the radial evolution of the outer shock in the equatorial plane as a function of time for a few of the runs with suffix "e1", corresponding to supernova energies of 1 FOE. Figs. 2b, 2c and 2d show the time-evolution of the volumes of gas with $T > 3X10^5$ K, $T > 7.8X10^5$ K and $T > 2.2X10^6$ K, respectively, for the same runs. These are the approximate temperatures at which the abundances of O VI, O VII, and O VIII peak in collisional ionization equilibrium (e.g., Gnat et al. 2007). Fig. 3a is analogous to Fig. 2a for runs with suffix "e2", corresponding to supernova energies of 2 FOE. Figs. 3b, 3c and 3d are analogous to Figs. 2b, 2c and 2d respectively and show the evolution of the volume of hot gas for runs with suffix "e2". We chose runs L0e1, L3e1, H6e1, VH20e1, L0e2, L3e2, H6e2 and VH20e2 for detailed analysis in this and the next few sections because they are representative of a low-density, high-density and very high-density ISM. Runs L0e1 and L3e1 represent unmagnetized and magnetized models with all other parameters kept constant. The same is true for runs L0e2 and L3e2. Such runs enable us to understand the specific differences introduced by the presence of a magnetic field.

Fig. 2a and 3a show us that the radial evolution of the outer shock does not vary significantly between the analogous simulations in these two figures. This is because the SNR spends a significant portion of its lifetime in the snow plow stage and the rate of expansion of the outer shock is weakly dependent on the energy of the explosion and strongly dependent on ambient density in this phase. Figs. 2b and 3b however show us that the volume of gas with $T > 3X10^5$ K is strongly dependent on the energy of the explosion and is consistent with the fact that stronger explosions produce larger bubbles of hot gas. Figs. 2c, 2d, 3c and 3d show similar trends with increasing explosion energy. We also see from Figs. 2b and 3b that the longevity of the hot gas bubbles decreases dramatically with increasing density. Figs. 2c, 2d, 3c and 3d show similar trends. As a result, we expect the four-volume of hot gas to have a strong negative correlation with interstellar density.

Comparing the L0e1 and L3e1 runs in Fig. 2b we see that the unmagnetized SNR has its $T > 3X10^5$ K gas refreshed at a time of 1.04 Myr. This time corresponds to the epoch when the hot gas bubble in L0e1 starts collapsing on itself. The gas with $T > 3 \times 10^5$ K gets refreshed in run L0e1 in response to the reverse shocks driven by the recollapsing walls of the bubble into the bubble cavity. From Fig. 1a we see that the thermal energy of the bubble in L3e1 is greater than that in the bubble of L0e1 because the heat flux out of the bubble in L3e1 is reduced by the anisotropic thermal conduction along the magnetic field. Fig. 1b shows us that the total thermal and magnetic energy in the shocked ISM that confines the bubble is much higher in L3e1. Consequently, the shocks that are driven into the hot gas bubbles in the magnetized runs tend to be much more vigorous due to the energy stored in the magnetic tension. As a result, the soft x-ray emitting gas in Fig. 2b shows a continuous variation with time. Similar trends can be seen between runs L0e2 and L3e2 in Fig. 3b. The evolution of the volumes of hot gas with $T > 7.8 \times 10^5$ K and $T > 2.2 \times 10^6$ K displayed in Fig. 2c, 2d, 3c, and 3d shows us that spikes appear in many of the runs at late epochs. This phenomenon is also caused by reheating of the gas by reverse shocks that propagate through the hot gas bubble as it recollapses. In all those instances we observe that the unmagnetized runs do not show reheating of gas with $T > 7.8 \times 10^5$ K during the collapse phase, while the magnetized runs do show this effect.

**III.2) Soft and Hard x-ray Fluxes**

Figs. 4a and 4b show the time-evolution of the soft (300-800 eV) and hard (1-5 keV) x-ray luminosities respectively for the runs L0e1, L3e1, H6e1 and VH20e1. Figs. 4c and 4d are analogous to Figs. 4a and 4b respectively but pertain to the runs L0e2, L3e2, H6e2 and VH20e2. We see that the SNRs emit in both soft and hard x-rays for a shorter duration as the ISM density increases. This shows us that x-ray luminosity tracks the lifetime of remnants. Table 2 catalogues the time taken for the hot gas bubble's collapse as observed in the simulations in the column marked $t_{collapse,plot}$. In Table 2 we also list $R_{max}$, the maximal equatorial radius of the hot gas bubble and $t_{collapse} = R_{max} / v_{ism}$. Fig. 4 can also be used to deduce the time it takes for the hot gas bubble to recollapse on itself. We see from Table 2 that $t_{collapse,plot}$ correlates very well with the time it takes for the soft and hard x-ray luminosities to fall to zero in Fig. 4. Table 2 also shows us that when the interstellar density is held constant, the increasing magnetic field usually results in a correlated decrease in $t_{collapse,plot}$ and $t_{collapse}$. The only loss of correlation between $t_{collapse,plot}$ and $t_{collapse}$ in Table 2 occurs for runs H20e1 and H20e2, which are strongly magnetized. Runs H20e1 and H20e2 are the only runs in our suite of runs that have a ratio of thermal pressure to magnetic pressure that is smaller than 1.5. As a result, the expansion in those two runs predominantly squeezes the ambient magnetic field, and the stored magnetic energy cannot be radiated away. The magnetic energy is retrieved during bubble collapse, giving those runs a longer lifetime.

The dominant trend seen in Figs. 4a and 4c is that the soft x-ray luminosity decreases almost monotonically with time. This is because soft x-ray emitting gas is abundantly present in the hot gas bubble and its thermal pressure is primarily responsible for the pressure support of the bubble. Comparing runs L0e1 and L3e1 in Fig. 4a shows us that the existence of a magnetic field plays an important role in maintaining the soft x-ray-emitting gas at late epochs in the SNR's evolution. This trend is also mirrored in Fig. 2b for the $T > 3 \times 10^5$ K gas. Similar trends can be found by comparing runs L0e2 and L3e2 in Fig. 4c.

We see in Figs. 4b and 4d that peaks appear in the hard x-ray luminosities at later epochs. The spikes in Figs. 4b and 4d mirror those in Figs. 2c and 3c for the gas with $T > 7.8 \times 10^5$ K showing that the hard x-rays correlate well with the gas at higher temperatures. These peaks are most prominent in the L3e1 and L3e2 runs, whereas the peaks are less pronounced in the H6e1 and H6e2 runs and almost nonexistent in the VH20e1 and VH20e2 runs. Table 2 shows that the runs with lower density have a greater maximum radius than the runs with higher density. Therefore, the runs with lower density compress the ambient magnetic field much more during the expansion phase, and the increased magnetic energy from this compression is available to reheat the gas as the bubble collapses. Furthermore, the energy contained in the compressed magnetic field is sufficient to reheat a small fraction of the gas in the bubble to temperatures where it radiates hard x-rays instead of soft x-rays.

**III.3) Quantifying the Amount of Hot Gas**

The filling factor of gas at various characteristic temperatures is an important quantity that bears on the interpretation of high ion and x-ray observations. Appropriate temperatures are those for which high-stage ions are strongly radiative. It is, therefore, interesting to quantify the porosity of hot gas, defined similarly to SC93 as $q_T = S_{-13} \int V_T(t)\, dt$ where $S_{-13}$ is the Galactic supernova rate in units of $10^{-13}$ pc$^{-3}$ yr$^{-1}$ and $V_T(t)$ is the volume of hot gas with temperature greater than "T" at a time "t" after the explosion. Specifically, we are interested in $q_{3\times10^5}$, $q_{7.8\times10^5}$ and $q_{2.2\times10^6}$ which correspond to gas with $T > 3\times10^5$, $T > 7.8\times10^5$ and $T > 2.2\times10^6$ which are temperatures at which OVI, OVII and OVIII begin to radiate. The porosity is a measure of the hot gas filling fraction, especially in situations where the supernova rate is not too high and the SNR shells do not intersect with each other. Figs. 2b, 2c, 2d, 3b, 3c, and 3d prove to be very useful for measuring the volume $V_T(t)$ of the hot gas at $T > 3\times10^5$ K, $T > 7.8\times10^5$ K, and $T > 2.2\times10^6$ K. Integrating the area under the curves in these figures, i.e. evaluating $\int V_T(t)\, dt$, also yields the four-volume. The four-volumes for all our runs have been catalogued in Tables 2 and 3.

Another important quantity affecting high-stage ion observations is the mean free path between encounters with SNR bubbles along select lines of sight. As shown by SC93, this is given by $\lambda_T = \left( S_{-13} \int A_T(t)\, dt \right)^{-1}$ where $A_T(t)$ is the area of hot gas with temperature greater than "T" at a time "t" after the explosion. We factor out $S_{-13}$ and show the results of the porosity as well as the reciprocal of the mean free path in Tables 2 and 3.

The dependence of the hot gas filling factors on the energy of the explosion $E_{51}$, the Galactic supernova rate $S_{-13}$, ISM atomic number density $n_0 = \rho_{ism}/(\mu m_p)$, ISM pressure $\tilde{P}_{04}\left(\equiv \left[B_{ism}^2/8\pi + R\rho_{ism}T_{ism}/\mu\right]/\left[k\,10^4\right]\right)$ are given below. ( $B_{ism}$, $\rho_{ism}$ and $T_{ism}$ in the previous formula for $\tilde{P}_{04}$ in cgs units.) Thus, we have used our simulations to parameterize $q_{3 \times 10^5}$, $q_{7.8 \times 10^5}$ and $q_{2.2 \times 10^6}$. We also do the same for $\lambda_{3 \times 10^5}$, $\lambda_{7.8 \times 10^5}$ and $\lambda_{2.2 \times 10^6}$. Consequently we have:

$$q_{3 \times 10^5} \approx 0.0629 S_{-13} E_{51}^{1.55} n_0^{-1.12} \left(\tilde{P}_{04}\right)^{-0.53} \tag{1}$$

$$q_{7.8 \times 10^5} \approx 0.0071 S_{-13} E_{51}^{1.76} n_0^{-2.03} \left(\tilde{P}_{04}\right)^{0.99} \tag{2}$$

$$q_{2.2 \times 10^6} \approx 0.000199 S_{-13} E_{51}^{2.04} n_0^{-1.68} \left(\tilde{P}_{04}\right)^{0.76} \tag{3}$$

$$\lambda_{3 \times 10^5}^{-1} \approx 2.42 \times 10^{10} \; S_{-13} \; \text{pc}^2 \; \text{yr} \left(E_{51}\right)^{1.21} \left(n_0\right)^{-1.26} \left(\tilde{P}_{04}\right)^{-0.027} \tag{4}$$

$$\lambda_{7.8 \times 10^5}^{-1} \approx 0.393 \times 10^{10} \; S_{-13} \; \text{pc}^2 \; \text{yr} \left(E_{51}\right)^{1.32} \left(n_0\right)^{-2.00} \left(\tilde{P}_{04}\right)^{1.23} \tag{5}$$

$$\lambda_{2.2 \times 10^6}^{-1} \approx 0.0154 \times 10^{10} \; S_{-13} \; \text{pc}^2 \; \text{yr} \left(E_{51}\right)^{1.58} \left(n_0\right)^{-1.55} \left(\tilde{P}_{04}\right)^{0.96} \tag{6}$$

The porosity and mean free path of the hot gas bubble at $T > 3 \times 10^5$ are also presented in the path-breaking work of SC93 as:

$$q_{3 \times 10^5} \approx 0.176 S_{-13} E_{51}^{1.17} n_0^{-0.61} \left(\tilde{P}_{04}\right)^{-1.06} \tag{7}$$

$$\lambda_{3 \times 10^5}^{-1} \approx 2.8 \times 10^{10} \; S_{-13} \; \text{pc}^2 \; \text{yr} \left(E_{51}\right)^{0.85} \left(n_0\right)^{-0.45} \left(\tilde{P}_{04}\right)^{-0.84} \tag{8}$$

We see that there is a reasonable concordance between eqns. (1) and (7) and then again between eqns. (4) and (8). It is, nevertheless, very interesting to try and understand the points of similarity as well as the differences between our results and SC93 because the differences illustrate the consequences of including different physics. Both codes can, of course, reproduce the results from CMB with high precision showing that for the same input physics the answers are indeed the same. The runs that SC93 report on do have lower explosion energies and lower densities on average. Most of the runs reported in SC93 are magnetically dominated. In contrast, in keeping with more recent measurements of interstellar magnetic fields (Beck et al 1999), a majority of our simulations are thermally dominated. Our 2.5d simulations include magnetic tension forces, while SC93 only include magnetic pressure forces. Our TC is anisotropic in the magnetized runs that we present, while SC93 used an isotropic formulation of TC in all their runs. SC93 also used non-equilibrium ionization and cooling while we use equilibrium cooling.

The coefficient in front of eqn. (1) is roughly concordant with that in front of eqn. (7) and similarly for eqns. (4) and (8). This reassures us that the same trends are being picked up by the singular value decomposition fitting procedure that we use to analyze our data (Press et al 1992). The parameters for our VL6e1 and VL6e2 runs are closest to similar runs reported in SC93 and for those runs, our $t_{collapse}$, $R_{max,bubble}$ and four-volume have values that agree quite well with SC93. The reason our coefficient in $q_{3x10^5}$ does not match up exactly with that in SC93 is because we use higher interstellar densities in our simulations. As a result, our simulations cool with greater radiative efficiency, thus robbing the expanding bubble of its energy and, consequently, reducing the four-volume.

We see that when $E_{51}$, $n_0$ and $\tilde{P}_{04}$ are set to unity eqns. (4) and (8) have very similar values, showing that both simulations produce the same values for the four-areas. This implies that both our models as well as SC93's models will produce the same projected area on the sky in a time-averaged sense. We see, however, that our coefficient in front of eqn. (1) is nearly three times smaller than that of SC93 in eqn. (7). This can be understood by realizing that the presence of magnetic tension in our simulations produces SNR morphologies that are elongated in the direction of the mean magnetic field. As a result, for the same average projected area, we expect bubbles that depart from sphericity to produce smaller amounts of four volume. This enables us to understand the differences in the coefficients in front of eqns. (1) and (7). We also observe that the dependence on explosion energy, $E_{51}$, is similar in eqns. (1) and (7) and then again in eqns. (4) and (8).

Many of our runs are dominated by thermal pressure whereas all of the runs in SC93 are dominated by magnetic pressure. (This is in large part due to the quite high thermal pressures for some of our simulations compared with theirs.) This makes it slightly risky to cross-compare density and pressure dependencies across eqns. (1) and

(7) and then again across eqns. (4) and (8). However, it is interesting to do so because it yields important insights, hence we do it here. The best perspective is had by writing $\tilde{P}_{04} \approx 10^{-4} \, n_0 \, T_{ism}$ in eqns. (1), (4), (7) and (8). On doing so we find that the variation of the porosities in eqns. (1) and (7) with density $n_0$ is indeed quite similar. Similarly, the mean free paths reported in eqns. (4) and (8) also show similar variation with density $n_0$. This agreement is to be expected because the density plays a major role in the evolution of both the Sedov and snowplow phases of an SNR. The temperature dependence in our eqn. (1) is given by $q_{3\times10^5} \propto T_{ism}^{-0.53}$ whereas eqn. (7) from SC93 yields $q_{3\times10^5} \propto T_{ism}^{-1.06}$. Thus SC93 find that their porosity has a stronger dependence on the temperature than we find in our simulations. This can be explained by realizing that the flux of thermal conduction is isotropic in SC93, whereas it is anisotropic for most of our runs. The total flux of thermal conduction in our simulations is, therefore, smaller in our simulations than in SC93, accounting for the weaker temperature dependence in our eqn. (1).

Comparing eqns. (2) and (3) to eqn. (1) shows that progressively hotter gas has smaller porosities, as expected. A similar trend can be seen by comparing eqns. (5) and (6) to eqn. (4). It is, however, very interesting to note that eqns. (2) and (3) have a positive correlation with increasing interstellar pressure $\tilde{P}_{04}$, whereas eqn. (1) shows the expected inverse correlation with interstellar pressure. This trend can be most easily understood by looking at Figs. 2b, 2c and 2d. Focusing on the model L3e1 we see that the gas with $T > 3 \times 10^5$ K survives for $\sim 10$ Myr while the gas with $T > 7.8 \times 10^5$ K almost disappears at 1.02 Myr only to reappear at 1.14 Myr. Fig. 2c further shows that once it reappears it is intermittently reheated till 6.35 Myr by the reverse shocks during the hot bubble's collapse. The H6e1 run in Fig. 2c shows a similar trend. In each case, we see that the volume of very hot gas generated during the collapse-induced reheating phase is indeed comparable to the volume of very hot gas during the initial explosion. Taking the run L3e1 in Fig. 2c as an example, we find that the four-volume after 1.14 Myr is 1.96 times larger than the four-volume before 1.02 Myr. The strength of the shocks formed during the bubble's collapse correlates positively with the interstellar pressure $\tilde{P}_{04}$. This explains the positive correlation between the porosity of very hot gas and $\tilde{P}_{04}$ reported in eqns. (2), (3), (5) and (6).

MO provide the following expression for the porosity:

$$q_{3\times 10^5} \approx 0.5\, S_{-13} E_{51}^{1.28} n_0^{-0.14} \left(\tilde{P}_{04}\right)^{-1.30}. \tag{9}$$

The porosity evaluated using eqn. (9) has been tabulated in Table 2. We see that their expression consistently gives an overestimate for the hot gas fraction by almost a factor of ten, a point also noted in SC93. SC93 used non-equilibrium ionization while we used equilibrium ionization in our simulations, and yet our two estimates agree on average while showing a significant departure from the estimate in MO. This shows us that it is not the details of ionization balance that regulates the hot gas fraction. The inclusion of thermal conduction is, therefore, seen to be the most important controlling factor in determining the fraction of hot gas. Our results for the hot gas fraction are, however, different from SC93 by a little more than a factor of two and we attribute that difference to our more careful treatment of magnetic fields and anisotropic thermal conduction as well as SC93's more careful treatment of the non-equilibrium ionization.

**III.4) Dosages and Line Widths of High-Stage Ions**

CHANDRA and FUSE observations of high-stage ions have recently become important in tracing the hot gas component in the ISM. The present calculations are all based on equilibrium ionization and cooling. As a result, a thorough study of high-stage ion densities from multi-dimensional simulations should await a non-equilibrium treatment. However, when that becomes available, it will be interesting to cross-compare with the equilibrium results presented here. For that reason, we present dosages, luminosities, and line widths of OVI, OVII and OVIII evaluated from our simulations. Following SC93, the definition of a dosage of a high stage ion "i" is given by

$$N_i(t) = \int n_i(t, \mathbf{r})\, dV \quad ; D_i = \int N_i(t)\, dt \tag{10}$$

where $n_i(t, \mathbf{r})$ is the local ion density at time "t" and position "$\mathbf{r}$". The dosages of OVI, OVII and OVIII are shown in Table 4. The line widths of the high-stage ions are also an observable quantity of interest. Figs. 5a and 5b show us the line widths and luminosities, respectively, of OVI for the L0e1, L3e1, H6e1, and VH20e1 simulations. Figs. 5c and 5d are analogous to Figs. 5a and 5b, respectively, except that they display data from the L0e2, L3e2, H6e2, and VH20e2 runs. OVI is a very good indicator of the presence of transitional temperatures where the SNR's shell begins to radiate. We see an absolute maximum for each simulation in Figs. 5b and 5d, which we conclude corresponds to the formation of the radiative snowplow stage. This data also confirms the trend that SNRs in denser mediums will have a shorter lifetime. This is due to the additional mass accumulating in front of the SNR that will result in an earlier snow plow stage and dispersion. Fig. 5a also shows that denser environments produce hotter bubbles early on in the evolution of the remnants. Consequently, the thermally broadened line widths for runs H6e1, and VH20e1 peak at earlier epochs and have higher line widths than runs L0e1 and L3e1. We see similar trends in Fig. 5c. The line widths in the later stages of all the remnants is ~ 30 Km/sec, in agreement with the results of SC92. The line shapes are consistent with Gaussian distributions.

Figs. 6a and 6b show us the line widths and luminosities, respectively, of OVII for the L0e1, L3e1, H6e1, and VH20e1 simulations. Figs. 6c and 6d are analogous to Figs. 6a and 6b, respectively, except that they display data from the L0e2, L3e2, H6e2, and VH20e2 runs. OVII is indicative of significantly hotter gas than OVI indicates, and therefore the graphs of its time-evolution do not track the onset of the radiative snowplow stage as well as the graphs of OVI. Figs. 7a and 7b show us the line widths and luminosities, respectively, of OVIII for the L0e1, L3e1, H6e1, and VH20e1 simulations. Figs. 7c and 7d are analogous to Figs. 7a and 7b, respectively, except that they display data from the L0e2, L3e2, H6e2, and VH20e2 runs. OVIII is effective in tracking the very hot gas in a bubble due to its high characteristic temperatures. We also see several spikes in Figs. 6b, 6d, 7b and 7d at later epochs, especially for the L3 simulations. This trend is similar to the graphs of soft x-ray and hard x-ray luminosities in Fig. 4, suggesting that the spikes are caused by the reverse shocks that are driven into the hot gas bubble by the compressed ambient magnetic field during the bubble's collapse. Similar spikes are not seen in Fig. 5 because OVI emitting gas is persistently present during the bubble's entire evolution. As a result, shocks that appear during the recollapse phase are not needed to produce OVI emitting gas. As we saw in Fig. 4, lower density runs have a greater $R_{max}$ value in Table 2, and consequently they compress the magnetic field to a greater extent.

The dependence of D(OVI), D(OVII) and D(OVIII) on the energy of the explosion $E_{51}$, the ISM atomic number density $n_0$, and the total ISM pressure $\tilde{P}_{04}$ are given below.

$$D(\text{O VI}) \approx 0.275 \times 10^{60} \text{ ion yr} (E_{51})^{0.56} (n_0)^{-1.12} (\tilde{P}_{04})^{-0.10} \tag{11}$$

$$D(\text{O VII}) \approx 3.31 \times 10^{60} \text{ ion yr} (E_{51})^{0.72} (n_0)^{-1.29} (\tilde{P}_{04})^{0.48} \tag{12}$$

$$D(\text{O VIII}) \approx 0.241 \times 10^{60} \text{ ion yr} (E_{51})^{0.91} (n_0)^{-0.63} (\tilde{P}_{04})^{0.18} \tag{13}$$

In comparison, SC93 derived a formula for the dosage of O VI that shows a greater dependency on the energy of the supernova:

$$D(\text{O VI}) \approx 8.13 \times 10^{60} \text{ ion yr} (E_{51})^{1.11} (n_0)^{-0.76} (\tilde{P}_{04})^{-0.16} \tag{14}$$

Since OVI is a species that sensitively depends non-equilibrium effects, we see that our equilibrium calculations in eqn. (11) yield an underestimate relative to SC93's eqn. (14). More detailed comparison requires an inclusion of the non-equilibrium effects in multi-dimensional calculations, an effect we will explore in a later paper.

**III.5) Line Widths of Radioactive Species**

The INTEGRAL satellite has recently measured the line profile of $^{26}$Al from various sources, Diehl et al (2006). $^{60}$Fe has also been measured and subsequent missions might even measure its line width. While $^{26}$Al can be ejected by the winds of massive stars as well as in SNe (Palacios et al 2005), it is quite likely that $^{60}$Fe is ejected exclusively from SNe (Timmes et al 1995) . The predicted ratios of $^{60}$Fe/$^{26}$Al depend strongly on high mass stellar evolution models and span a range of values due to variations in the models (Timmes et al 1995, Palacios et al 2005, Limongi & Chieffi 2006, Woosley & Heger 2007). For that reason, matching the observed $^{60}$Fe/$^{26}$Al ratio is not one of the goals of the present study. However, line widths of these two elements can be safely predicted based on SNR models and is a goal of this work. $^{26}$Al and $^{60}$Fe undergo inverse beta decay with lifetimes of 0.716 Myr and 1.49 Myr. Since the supernova ejecta would bear these radioactive elements, it is interesting to evaluate the line widths of radioactive species that are borne by SN ejecta. A more complete calculation would have to include the propagation of these radioactive species in the ISM and should include the decay of the ejecta as they propagate through the ISM. We defer such a study for later. In this paper we note that the lifetimes for these radioactive species is shorter than the dispersal times for SNRs, making it possible to invoke a simplification. We, therefore, ignore the SNR dispersal and present simulated line widths from the ejecta as a function of time.

Fig. 8a shows the line widths of $^{26}$Al as a function of time for the L0e1, L3e1, H6e1 and VH20e1 runs while Fig. 8b shows the same for $^{60}$Fe as a function of time for the same runs. Fig. 8c shows the line widths of $^{26}$Al as a function of time for the L0e2, L3e2, H6e2 and VH20e2 runs while Fig. 8d shows the same for $^{60}$Fe as a function of time for the same runs. Comparing Figs. 5a, 8a and 8b we see that the line widths of OVI, $^{26}$Al and $^{60}$Fe are progressively smaller at similar times in the same runs. This reflects the role of thermal broadening at early epochs and shows that atoms with higher atomic mass will show lesser thermal broadening. Similar to trends seen in Figs. 5a, 6a and 7a we see from Figs. 8a and 8b that remnants expanding into denser environments have broader line widths at early epochs in their evolution as well as a shorter duration over which these broader line widths present themselves.

Fig. 8 also shows us that the line widths from the individual remnants simulated here remain below ~ 200 Km/sec for much of the duration of the simulations. (This result holds for times greater than $10^4$ years where we produced simulated data that is well-sampled in time.) Furthermore, individual remnants expanding into low density environments have radioactive line widths that are consistently lower than 200 Km/sec. Remnants that expand into denser environments have hotter bubbles during their early evolutionary epochs and, therefore, show significantly higher thermal broadening of the line widths. INTEGRAL has found that the measured line widths of $^{26}$Al are quite possibly less than 340 Km/sec (Diehl 2006). More recent INTEGRAL data seems to constrain the $^{26}$Al line width to lie between several tens of Km/sec and ~ 150Km/sec (Diehl, 2007). Should these line widths arise in individual remnants, the upper bounds on the observed line widths in Diehl (2006) are consistent with our simulations. It is worth pointing out that the interiors of remnants would be hotter by an order of magnitude if thermal conduction had not been included (Tilley & Balsara 2006), resulting in anticipated line widths that are substantially larger than the ones predicted here. This probably explains why the large line widths that were deduced from gamma-ray observations that preceeded INTEGRAL were not held to be very surprising. It is also worth pointing out that many of the environments targeted by INTEGRAL's observations tend to be star forming regions which tend to be denser parts of the ISM. Fig. 8 shows that for such high density regions individual remnants spend a relatively small time in a mode where their radioactive line widths would exceed a couple of hundred Km/sec. Consequently, the fact that such targeted observations do not produce large observed line widths for $^{26}$Al is not surprising in light of our simulations.

Comparing Figs. 8a to 8b and also comparing Figs. 8c and 8d we observe another interesting trend. We see that in the early epoch of any given remnant's evolution the line widths of $^{60}$Fe are substantially smaller than the line widths of $^{26}$Al, while that trend disappears as the remnants evolve. While INTEGRAL cannot measure the line widths of $^{60}$Fe, this trend has an interesting observable consequence. Observations showing $^{26}$Al line widths that significantly exceed those of $^{60}$Fe would be indicative of young SNRs. Observations of similar line widths would, however, originate from older remnants.

**IV) Discussion**

The study of SNRs presented in this paper shows that the late stages in the evolution of remnants play an important role in determining observationally relevant quantities such as the four volumes of hot gas, the luminosities and line widths of various high-stage ions and the line widths of radioactive species. It is commonly thought that the remnant undergoes turbulent dispersal at the end of the snowplow phase. However, characterizing the time for the onset of turbulent dispersal and the timescale over which it acts has been difficult. Direct simulations of a SN-driven turbulent ISM, see Fig. 2 from Mac Low et al (2005), show that an outline of the remnant persists well after a few million years. Such simulations make it plausible that the hot gas bubble may exist as a coherent entity for a few Myr, even as interstellar turbulence gradually diffuses into it. Table 2 shows that the hot gas bubbles in our simulations collapse upon themselves in less than 21 Myr in all cases, with a collapse time between 2 and 7 Myr being very common. Table 2 also shows that a majority of the remnants have radii given by $R_{max,bubble}$ that lies between 15 and 60 pc. We thus assume a characteristic hot gas bubble diameter of 100 pc and a characteristic collapse time for the bubble of ~5 Myr. As a result, if we can demonstrate that interstellar turbulence cannot encroach upon a region of 100 pc in less than 5 Myrs then we can safely assert that the late stages of SNR evolution computed in this paper will indeed contribute to the observables as predicted here.

A better argument for the persistence of SNR-driven hot gas bubbles may be constructed from studies of turbulent mixing (Bateman & Larson 1993, Roy & Kunth 1995, Oey 2003, Klessen & Lin 2003, Balsara & Kim 2005). Using a supernova rate that is 8 times the current Galactic rate, Balsara & Kim (2005) found a turbulent diffusion coefficient given by $5.7 \times 10^{26}$ cm$^2$ / sec. While this turbulent diffusion would be an over-estimate at the solar circle of our Galaxy, it would possibly be quite appropriate in the inner parts of our Galaxy. For our stated turbulent diffusion coefficient, we find that the turbulence would diffuse into a 100 pc region in ~ 5.3 Myrs. The supernova rate at our solar circle is less than the one used in Balsara & Kim (2005) yielding a smaller turbulent diffusion coefficient and a corresponding larger time for turbulent diffusion. We have, therefore, shown that the late stage results presented in this paper can indeed be observationally relevant.

**V) Conclusions**

We have carried out a large number of 2.5d simulations of SNRs evolving in interstellar media having a range of densities, temperatures and magnetic field strengths. Two different energies have been used for the SNe explosions. The simulations include the effects of equilibrium radiative cooling and anisotropic thermal conduction in the presence of magnetic fields. The simulations enable us to come to the following conclusions:

1) The volume of hot gas as well as its longevity is strongly dependent on the ISM density and weakly dependent on the explosion energy. Consequently, SNRs emit in both soft and hard x-rays for a shorter duration as the ISM density increases.

2) The presence and strength of the magnetic field plays a very important role in re-energizing the gas at later epochs in the hot bubble's collapse phase. Compression of the magnetic field plays an important role in storing a significant fraction of the remnant's expansion energy and this energy plays an important role in reheating the hot gas bubble during collapse. Low density magnetized runs produce the largest amount of magnetic compression and, therefore, show the greatest effect of reheating. Unmagnetized remnants show a much weaker reheating effect and are unable to reheat the hot gas bubble above $T > 7.8 \times 10^5$ K.

3) The time taken by the hot gas bubble to collapse has been catalogued in the paper and correlates well with $R_{max} / v_{ism}$, where $R_{max}$ is the bubble's maximal radial extent and $v_{ism}$ is the maximal signal speed in the ISM.

4) We have catalogued simulated four-volumes and three-areas for the gas with $T > 3 \times 10^5$, $T > 7.8 \times 10^5$ and $T > 2.2 \times 10^6$ which are temperatures at which OVI, OVII and OVIII begin to radiate. We have used these values to parametrize the dependence of porosity and mean free path between encounters with SNR bubbles as a function of explosion energy, ISM density, interstellar pressure and the Galactic SN rate. Our analysis of these results shows the importance of magnetic tension forces as well as the importance of anisotropic thermal conduction along field lines.

5) Our estimates for the hot gas content are closely concordant with those of SC92 and SC93. They show that thermal conduction causes a significant reduction in the four-volume estimated by MO.

6) We also catalogue the simulated luminosities and line widths of OVI, OVII and OVIII as well as their dosages. These have been obtained with the simplifying approximation of equilibrium ionization for eventual comparison with a non-equilibrium ionization calculation.

7) Simulated line widths of radioactive species, $^{26}$Al and $^{60}$Fe, ejected by SNe have also been catalogued. In many instances we find that these line widths are less than 200 Km/sec for most of the SNRs evolution. This enables us to understand why INTEGRAL has not been able to detect very large line widths in certain star forming regions. The importance of thermal conduction in obtaining smaller simulated values for the line widths of radioactive species has also been catalogued.

8) We find that in the early epoch of any given remnant's evolution the line widths of $^{60}$Fe are substantially smaller than the line widths of $^{26}$Al, while that trend disappears as the remnants evolve. Future observations could be in a position to confirm this prediction.

**Acknowledgements**

We acknowledge support via NSF-PFC grant PHY02-16783, NSF grant AST-0607731 and NASA grant HST-AR-10934.01-A. The majority of simulations were performed on PC clusters at UND but a few initial simulations were also performed at NASA-NCCS.

**Figure Captions**

Fig. 1a shows the time evolution of the volume-weighted mean temperature of the hot gas bubble. The solid line corresponds to L0e1 and the dashed line corresponds to L3e1. Fig. 1b shows the time evolution of the thermal and magnetic energy in the hot gas bubbles as well as the same in the shocked ISM. The solid line in Fig. 1b shows the thermal and magnetic energy in the hot gas bubble of L0e1, the dashed line displays the same for L3e1. The dotted line in Fig. 1b shows the thermal and magnetic energy in the shocked interstellar gas for L0e1 while the dot-dashed line shows the same for L3e1.

Fig. 2a shows the radial evolution of the outer shock in the equatorial plane as a function of time for the simulations L0e1, L3e1, H6e1, and VH20e1, corresponding to supernova energies of 1 FOE. Figs. 2b, 2c and 2d show the time-evolution of the volumes of gas with $T > 3 \times 10^5$ K, $T > 7.8 \times 10^5$ K and $T > 2.2 \times 10^6$ K respectively, for the same simulations.

Fig. 3a is analogous to Fig. 2a for the simulations L0e2, L3e2, H6e2, and VH20e2, corresponding to supernova energies of 2 FOE. Figs. 3b, 3c and 3d are analogous to Figs. 2b, 2c and 2d, respectively, but they pertain to the same simulations as in Fig. 3a.

Figs. 4a and 4b show the time-evolution of the soft (300-800 eV) and hard (1-5 keV) x-ray luminosities, respectively, for the simulations labeled L0e1, L3e1, H6e1, and VH20e1. Figs. 4c and 4d are analogous to Figs. 4a and 4b, respectively, but pertain to the simulations labeled L0e2, L3e2, H6e2, and VH20e2.

Figs. 5a and 5b show line widths and luminosities, respectively, for the emission from OVI as a function of time for the simulations labeled L0e1, L3e1, H6e1, and VH20e1. Figs. 5c and 5d are analogous to Figs. 5a and 5b, respectively, except that they pertain to the simulations labeled L0e2, L3e2, H6e2, and VH20e2.

Figs. 6a and 6b show line widths and luminosities, respectively, for the emission from OVII as a function of time for the simulations labeled L0e1, L3e1, H6e1, and VH20e1. Figs. 6c and 6d are analogous to Figs. 6a and 6b, respectively, except that they pertain to the simulations labeled L0e2, L3e2, H6e2, and VH20e2.

Figs. 7a and 7b show line widths and luminosities, respectively, for the emission from OVIII as a function of time for the simulations labeled L0e1, L3e1, H6e1, and VH20e1. Figs. 7c and 7d are analogous to Figs. 7a and 7b, respectively, except that they pertain to the simulations labeled L0e2, L3e2, H6e2, and VH20e2.

Fig. 8a shows the time-evolution of the simulated line widths of radioactive $^{26}$Al for the simulations labeled L0e1, L3e1, H6e1, and VH20e1. Fig. 8b does the same for $^{60}$Fe. Figs. 8c and 8d are analogous to Figs. 8a and 8b respectively, except that they pertain to the simulations labeled L0e2, L3e2, H6e2, and VH20e2.

Table 1 shows the parameters for the simulations presented here.

| Run Name | $R_{outer}$ (pc) | $N_{zones}$ | $t_{final}$ (Myr) | $\rho_{ism}$ ($10^{-24}$ gm/cm$^3$) | $T_{ism}$ (Kelvin) | $B_{ism}$ (μGauss) | $v_{ism}$ (Km/sec) | FOE |
|---|---|---|---|---|---|---|---|---|
| VL0e1 | 400 | 384 | 30 | 0.2 | 10,000 | 0 | 15.36 | 1 |
| VL2e1 | 400 | 384 | 30 | 0.2 | 10,000 | 2 | 19.98 | 1 |
| VL6e1 | 400 | 384 | 30 | 0.2 | 10,000 | 6 | 41.30 | 1 |
| L0e1 | 300 | 384 | 9 | 0.7 | 8,000 | 0 | 13.74 | 1 |
| L3e1 | 300 | 384 | 9 | 0.7 | 8,000 | 3 | 17.14 | 1 |
| L6e1 | 300 | 384 | 9 | 0.7 | 8,000 | 6 | 24.67 | 1 |
| H0e1 | 100 | 192 | 4 | 5.0 | 10,000 | 0 | 15.36 | 1 |
| H6e1 | 100 | 192 | 4 | 5.0 | 10,000 | 6 | 17.17 | 1 |
| H20e1 | 100 | 192 | 4 | 5.0 | 10,000 | 20 | 29.82 | 1 |
| VH0e1 | 100 | 192 | 2 | 20.0 | 10,000 | 0 | 15.36 | 1 |
| VH6e1 | 100 | 192 | 2 | 20.0 | 10,000 | 6 | 15.83 | 1 |
| VH20e1 | 100 | 192 | 2 | 20.0 | 10,000 | 20 | 19.98 | 1 |
| VL0e2 | 400 | 384 | 30 | 0.2 | 10,000 | 0 | 15.36 | 2 |
| VL2e2 | 400 | 384 | 30 | 0.2 | 10,000 | 2 | 19.98 | 2 |
| VL6e2 | 400 | 384 | 30 | 0.2 | 10,000 | 6 | 41.30 | 2 |
| L0e2 | 300 | 384 | 9 | 0.7 | 8,000 | 0 | 13.74 | 2 |
| L3e2 | 300 | 384 | 9 | 0.7 | 8,000 | 3 | 17.14 | 2 |
| L6e2 | 300 | 384 | 9 | 0.7 | 8,000 | 6 | 24.67 | 2 |
| H0e2 | 100 | 192 | 4 | 5.0 | 10,000 | 0 | 15.36 | 2 |
| H6e2 | 100 | 192 | 4 | 5.0 | 10,000 | 6 | 17.17 | 2 |
| H20e2 | 100 | 192 | 4 | 5.0 | 10,000 | 20 | 29.82 | 2 |
| VH0e2 | 100 | 192 | 2 | 20.0 | 10,000 | 0 | 15.36 | 2 |
| VH6e2 | 100 | 192 | 2 | 20.0 | 10,000 | 6 | 15.83 | 2 |
| VH20e2 | 100 | 192 | 2 | 20.0 | 10,000 | 20 | 19.98 | 2 |

Table 2 shows observable quantities that are extracted from the simulations presented here.

| Run Name | $R_{max,bubble}$ (pc) | $t_{collapse}$(plot) (Myr) | $t_{collapse} = \frac{R_{max}}{v_{ism}}$ (Myr) | $\int V_{3\times10^5}(t)\,dt$ (pc$^3$ yr) | $q_{3\times10^5}/S_{-13}$ from MO (pc$^3$ yr) | $\int A_{3\times10^5}(t)\,dt$ (pc$^2$ yr) |
|---|---|---|---|---|---|---|
| VL0e1 | 93.47 | 14.8 | 5.95 | 7.65e12 | 2.49e13 | 1.62e11 |
| VL2e1 | 84.12 | 16.1 | 4.12 | 1.10e13 | 2.49e13 | 2.89e11 |
| VL6e1 | 66.47 | 13.27 | 1.57 | 3.89e12 | 2.49e13 | 2.01e11 |
| L0e1 | 59.20 | 6.99 | 4.21 | 8.92e11 | 5.47e12 | 2.95e10 |
| L3e1 | 52.97 | 6.35 | 3.02 | 1.00e12 | 5.47e12 | 3.79e10 |
| L6e1 | 46.73 | 5.82 | 1.85 | 6.89e11 | 5.47e12 | 3.52e10 |
| H0e1 | 28.56 | 2.68 | 1.82 | 5.59e10 | 2.41e11 | 3.45e9 |
| H6e1 | 24.92 | 1.95 | 1.42 | 3.05e10 | 2.41e11 | 2.42e9 |
| H20e1 | 20.77 | 3.35 | 0.68 | 2.43e10 | 2.41e11 | 2.81e9 |
| VH0e1 | 16.88 | 1.70 | 1.07 | 8.05e9 | 3.28e10 | 8.36e8 |
| VH6e1 | 15.58 | 1.41 | 0.96 | 5.24e9 | 3.28e10 | 6.40e8 |
| VH20e1 | 13.24 | 0.79 | 0.65 | 2.16e9 | 3.28e10 | 3.23e8 |
| VL0e2 | 118.39 | 20.1 | 7.54 | 2.45e13 | 6.03e13 | 3.92e11 |
| VL2e2 | 106.97 | 21.4 | 5.24 | 3.01e13 | 6.03e13 | 6.25e11 |
| VL6e2 | 83.08 | N/A | 1.97 | 1.11e13 | 6.03e13 | 5.30e11 |
| L0e2 | 74.77 | 9.18 | 5.32 | 2.68e12 | 1.33e13 | 6.79e10 |
| L3e2 | 65.43 | 8.42 | 3.73 | 2.67e12 | 1.33e13 | 8.01e10 |
| L6e2 | 58.42 | 7.67 | 2.32 | 1.76e12 | 1.33e13 | 7.08e10 |
| H0e2 | 36.09 | 4.80 | 2.30 | 2.12e11 | 5.86e11 | 1.05e10 |
| H6e2 | 32.45 | 3.86 | 1.85 | 1.45e11 | 5.86e11 | 8.37e9 |
| H20e2 | 25.70 | 5.21 | 0.84 | 6.51e10 | 5.86e11 | 6.05e9 |
| VH0e2 | 21.03 | 2.21 | 1.34 | 2.08e10 | 7.96e10 | 1.74e9 |
| VH6e2 | 18.95 | 1.78 | 1.17 | 1.30e10 | 7.96e10 | 1.25e9 |
| VH20e2 | 16.36 | 1.07 | 0.80 | 5.53e9 | 7.96e10 | 6.60e8 |

Table 3 shows four-volumes and three-areas of hot gas at various temperatures and dosages of OVI, OVII and OVIII that are extracted from the simulations presented here.

| Run Name | $\int V_{7.8\times10^5}(t)\,dt$ (pc$^3$ yr) | $\int V_{2.2\times10^6}(t)\,dt$ (pc$^3$ yr) | $\int A_{7.8\times10^5}(t)\,dt$ (pc$^2$ yr) | $\int A_{2.2\times10^6}(t)\,dt$ (pc$^2$ yr) |
|---|---|---|---|---|
| VL0e1 | 1.13e11 | 9.04e9 | 3.48e9 | 4.33e8 |
| VL2e1 | 2.61e12 | 2.67e10 | 1.23e11 | 1.33e9 |
| VL6e1 | 2.16e12 | 2.70e10 | 1.10e11 | 1.41e9 |
| L0e1 | 2.01e10 | 2.34e9 | 9.41e8 | 1.63e8 |
| L3e1 | 2.10e11 | 5.09e9 | 1.45e10 | 3.79e8 |
| L6e1 | 4.00e11 | 5.26e9 | 2.47e10 | 3.97e8 |
| H0e1 | 2.23e10 | 2.15e8 | 1.79e9 | 2.84e7 |
| H6e1 | 2.11e10 | 3.60e8 | 1.88e9 | 4.79e7 |
| H20e1 | 2.15e10 | 1.55e9 | 2.40e9 | 3.34e8 |
| VH0e1 | 6.03e9 | 3.47e7 | 6.68e8 | 7.04e6 |
| VH6e1 | 4.83e9 | 9.54e8 | 5.97e8 | 1.12e8 |
| VH20e1 | 2.00e9 | 8.05e7 | 3.06e8 | 2.90e7 |
| VL0e2 | 2.87e11 | 2.60e10 | 6.98e9 | 9.45e8 |
| VL2e2 | 1.29e13 | 8.32e10 | 3.75e11 | 3.16e9 |
| VL6e2 | 6.23e12 | 8.43e10 | 2.54e11 | 3.32e9 |
| L0e2 | 5.05e10 | 6.59e9 | 1.86e9 | 3.56e8 |
| L3e2 | 8.47e11 | 1.51e10 | 3.96e10 | 8.57e8 |
| L6e2 | 1.25e12 | 1.52e10 | 5.50e10 | 8.81e8 |
| H0e2 | 1.31e11 | 6.21e8 | 7.40e9 | 6.10e7 |
| H6e2 | 1.32e11 | 6.90e9 | 7.84e9 | 4.99e8 |
| H20e2 | 5.91e10 | 8.71e9 | 5.34e9 | 1.37e9 |
| VH0e2 | 1.72e10 | 8.99e7 | 1.50e9 | 1.41e7 |
| VH6e2 | 1.23e10 | 4.78e9 | 1.19e9 | 4.64e8 |
| VH20e2 | 5.27e9 | 6.68e8 | 6.36e8 | 1.40e8 |

Table 4 shows dosages of OVI, OVII and OVIII that are extracted from the simulations presented here.

| Run Name | D(OVI) (ion yr) | D(OVII) (ion yr) | D(OVIII) (ion yr) |
|---|---|---|---|
| VL0e1 | 2.82e60 | 5.60e60 | 4.47e59 |
| VL2e1 | 1.13e60 | 2.44e61 | 5.29e59 |
| VL6e1 | 1.99e60 | 3.10e61 | 5.63e59 |
| L0e1 | 5.76e59 | 2.74e60 | 3.23e59 |
| L3e1 | 2.17e59 | 5.41e60 | 3.34e59 |
| L6e1 | 3.29e59 | 8.64e60 | 3.28e59 |
| H0e1 | 5.69e58 | 1.57e60 | 1.63e59 |
| H6e1 | 3.09e58 | 1.11e60 | 1.37e59 |
| H20e1 | 2.80e58 | 9.24e59 | 1.71e59 |
| VH0e1 | 1.13e58 | 4.99e59 | 5.16e58 |
| VH6e1 | 8.98e57 | 2.99e59 | 6.13e58 |
| VH20e1 | 7.00e57 | 2.60e59 | 4.98e58 |
| VL0e2 | 4.92e60 | 1.02e61 | 8.48e59 |
| VL2e2 | 1.87e60 | 4.84e61 | 1.04e60 |
| VL6e2 | 4.70e60 | 6.06e61 | 1.14e60 |
| L0e2 | 9.94e59 | 4.65e60 | 6.43e59 |
| L3e2 | 3.85e59 | 1.11e61 | 5.87e59 |
| L6e2 | 5.14e59 | 1.52e61 | 6.10e59 |
| H0e2 | 5.09e58 | 2.20e60 | 1.44e59 |
| H6e2 | 2.87e58 | 1.36e60 | 1.75e59 |
| H20e2 | 5.15e58 | 1.62e60 | 5.07e59 |
| VH0e2 | 1.69e58 | 8.59e59 | 1.03e59 |
| VH6e2 | 1.05e58 | 3.95e59 | 1.46e59 |
| VH20e2 | 8.38e57 | 3.62e59 | 1.22e59 |

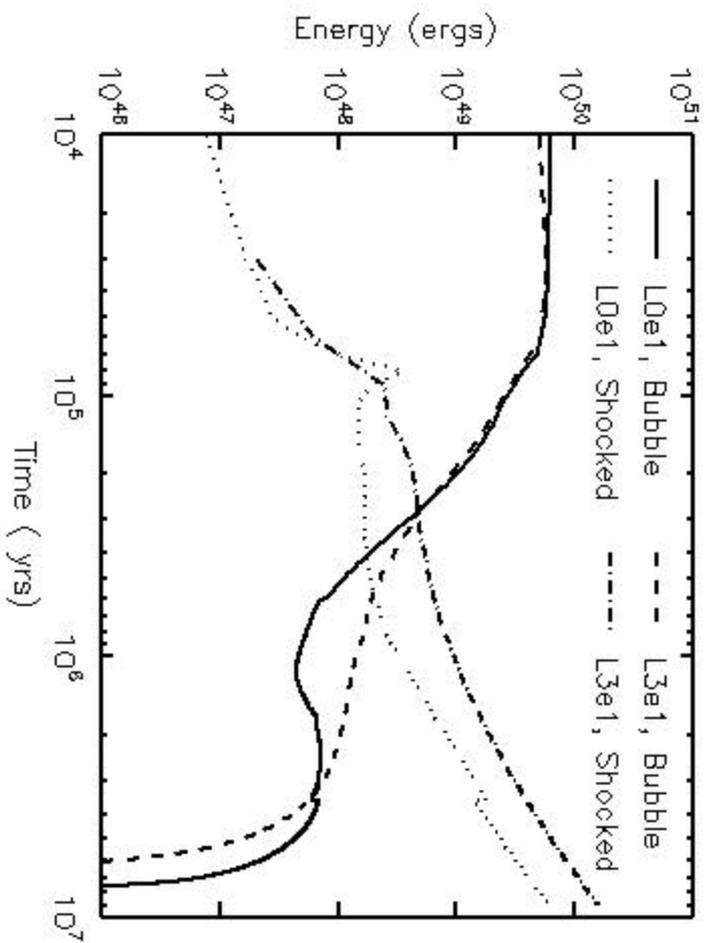

Fig. 1a

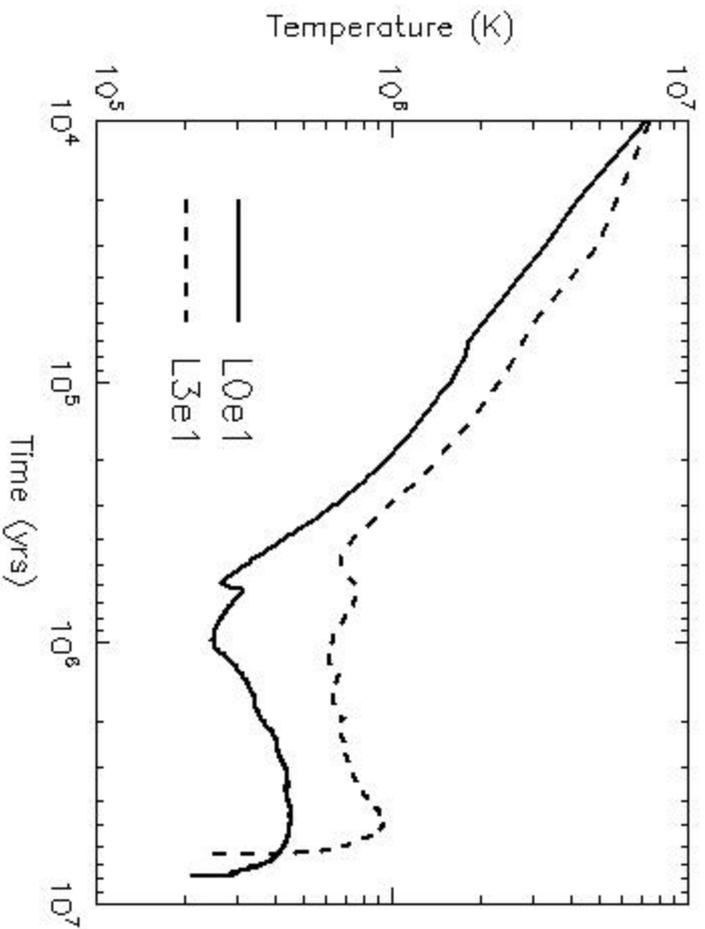

Fig. 1b

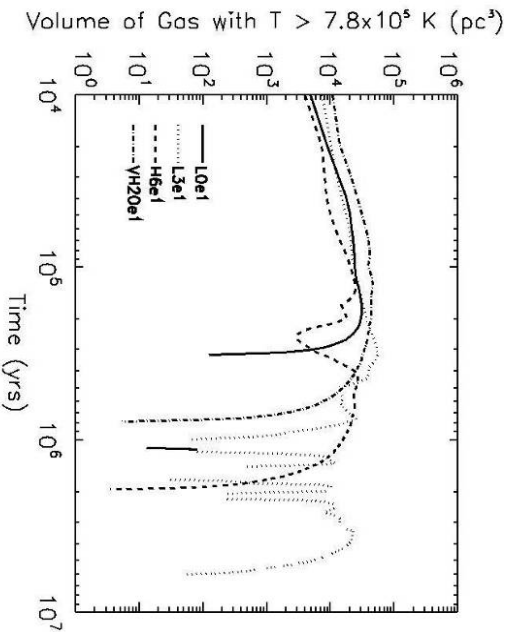

Fig. 2c

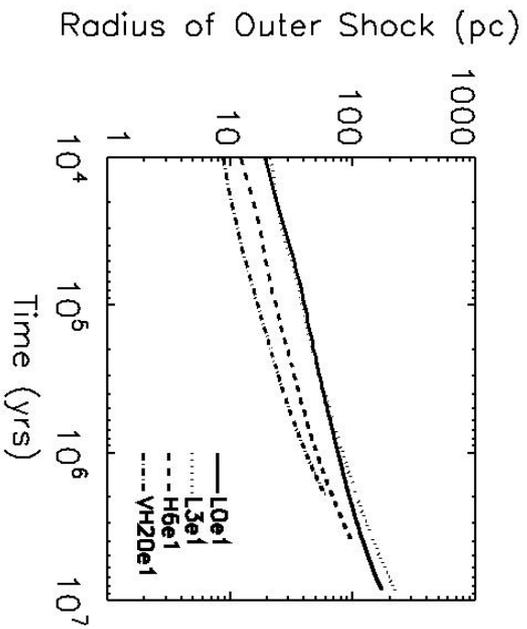

Fig. 2a

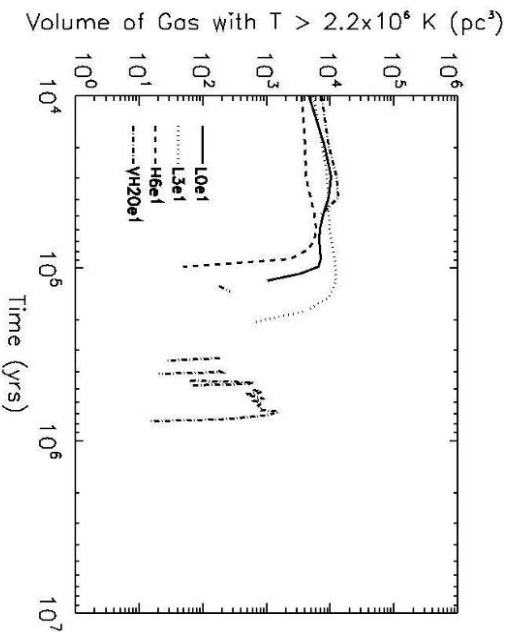

Fig. 2d

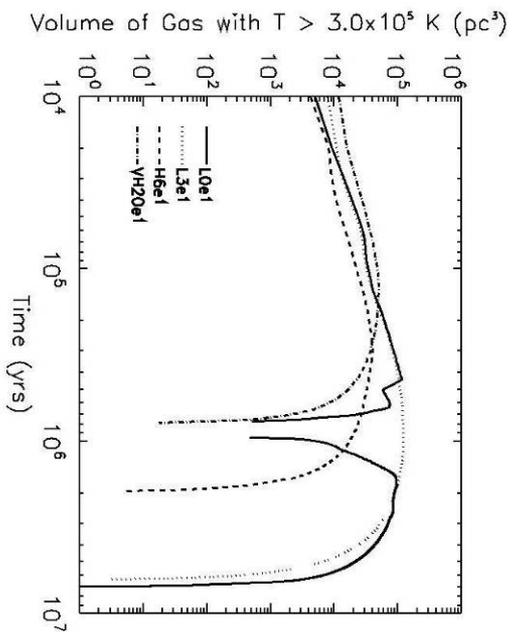

Fig. 2b

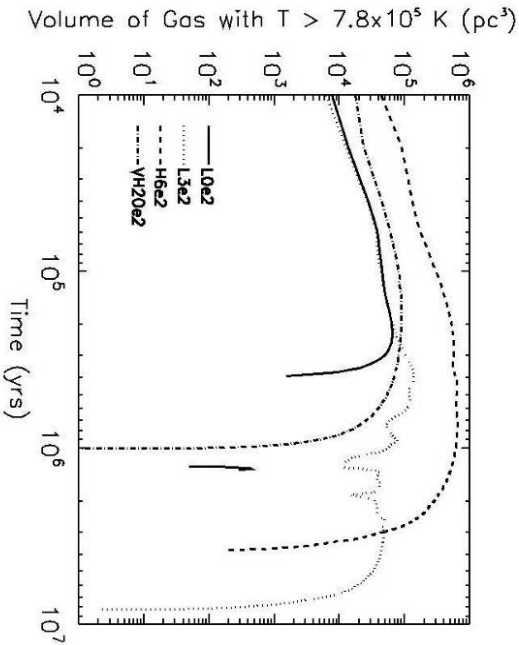

Fig. 3c

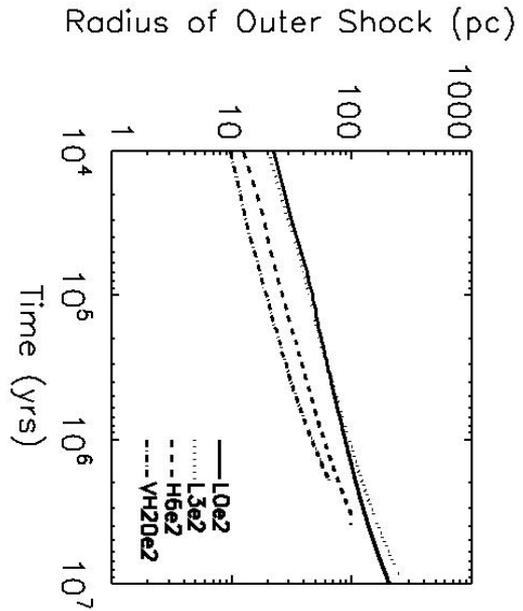

Fig. 3a

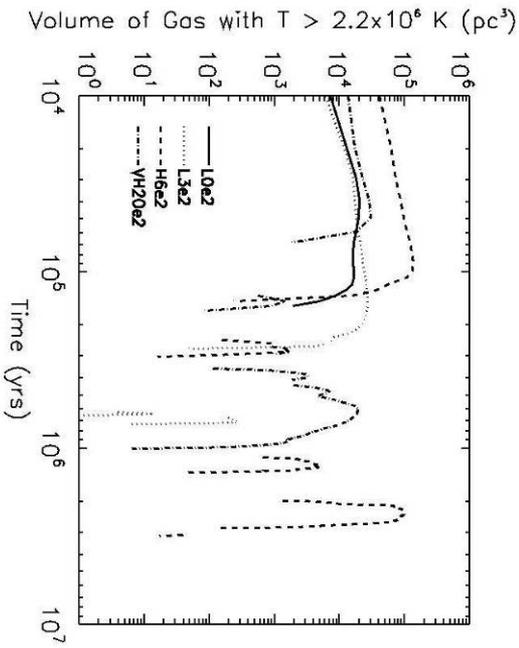

Fig. 3d

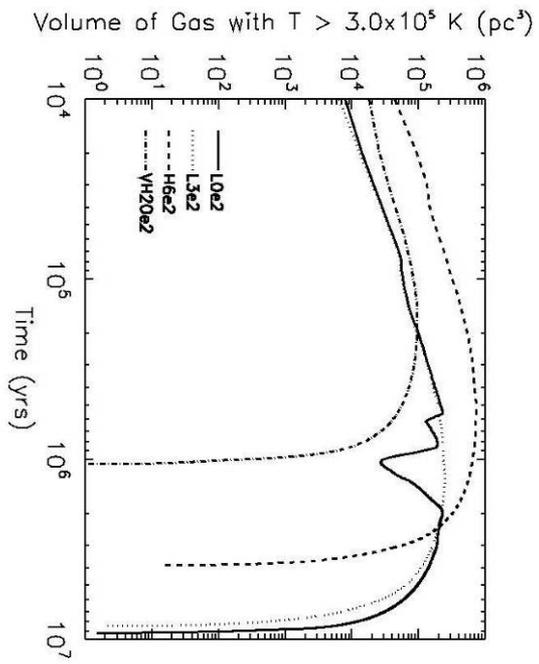

Fig. 3b

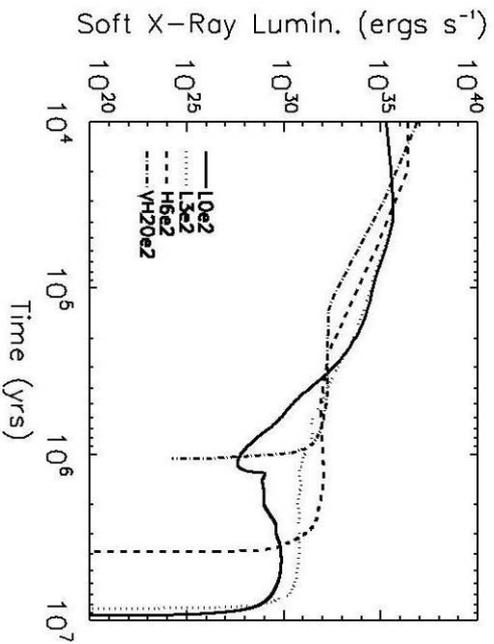

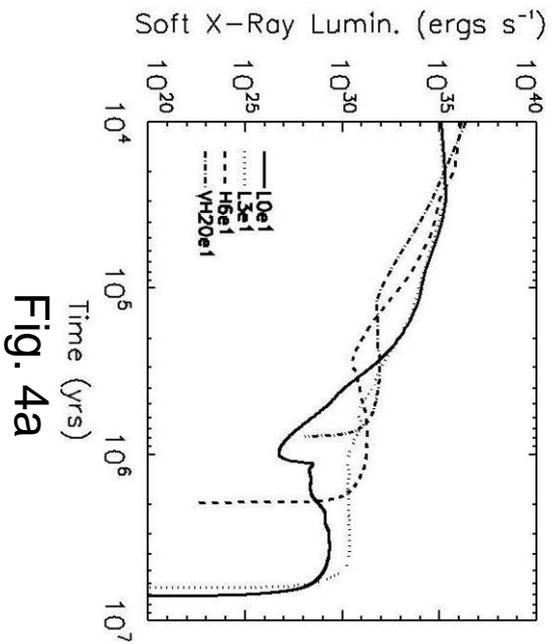

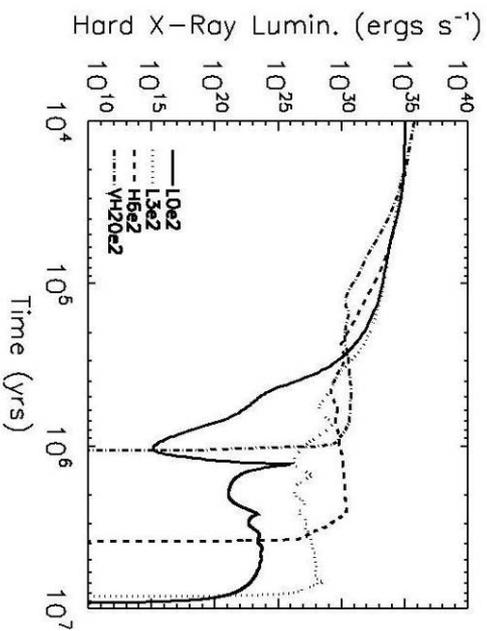

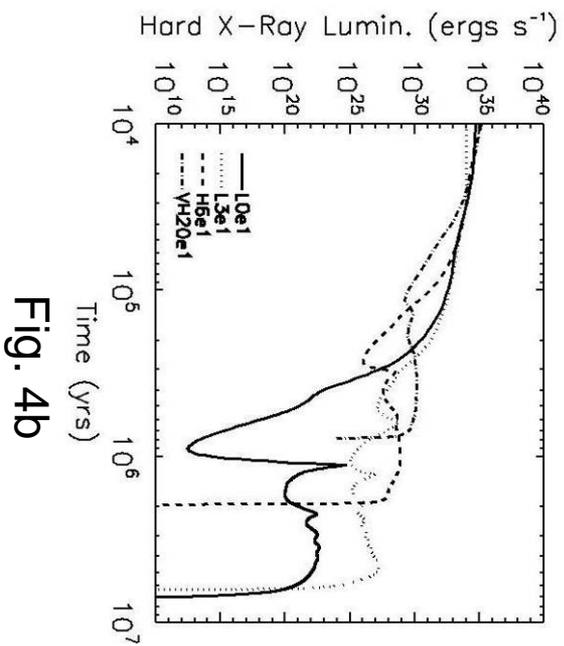

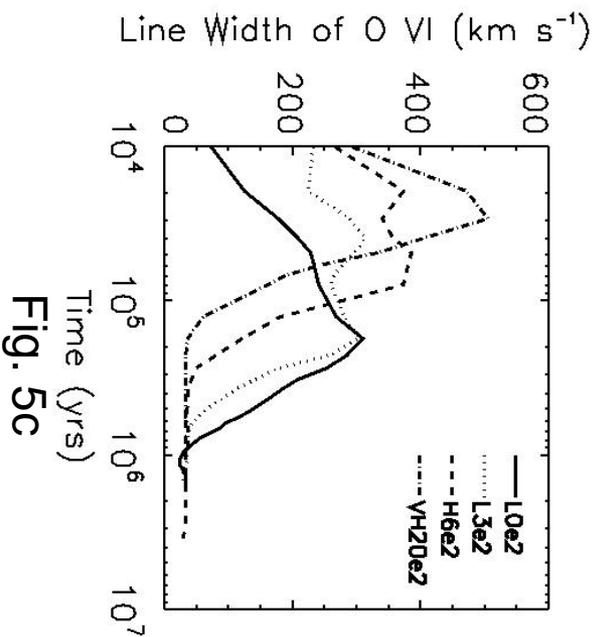

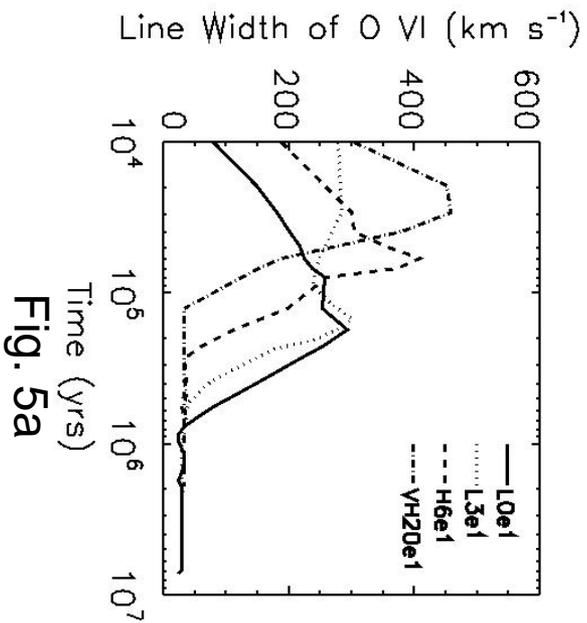

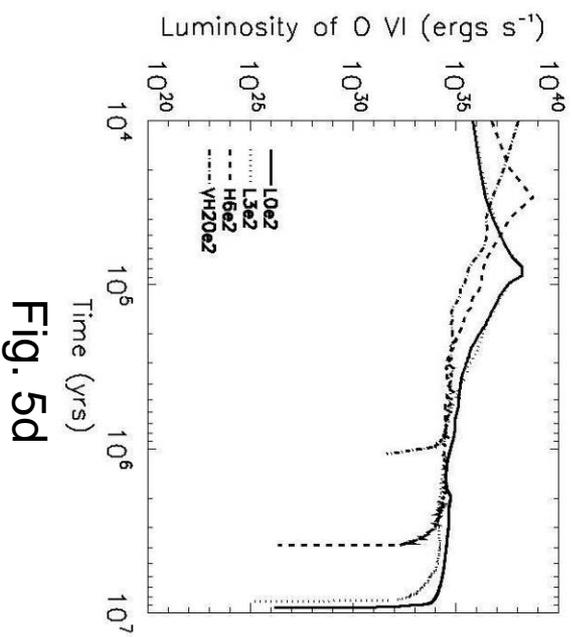

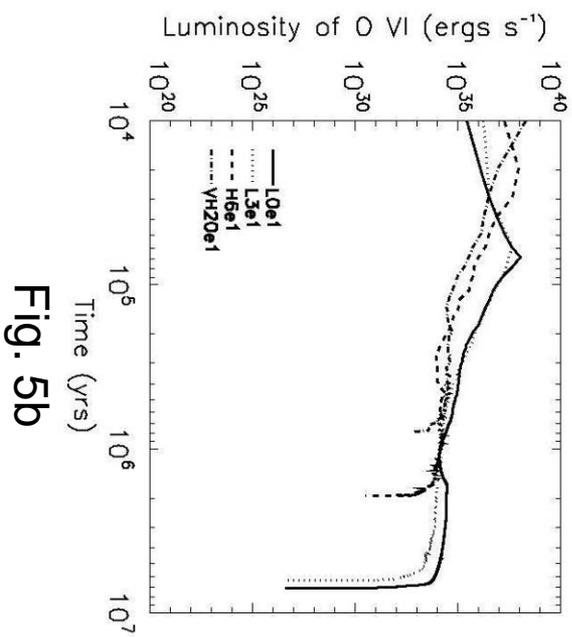

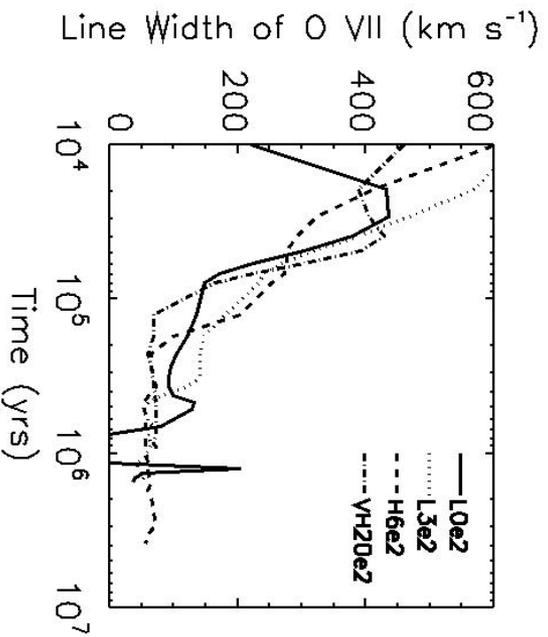

Fig. 6a

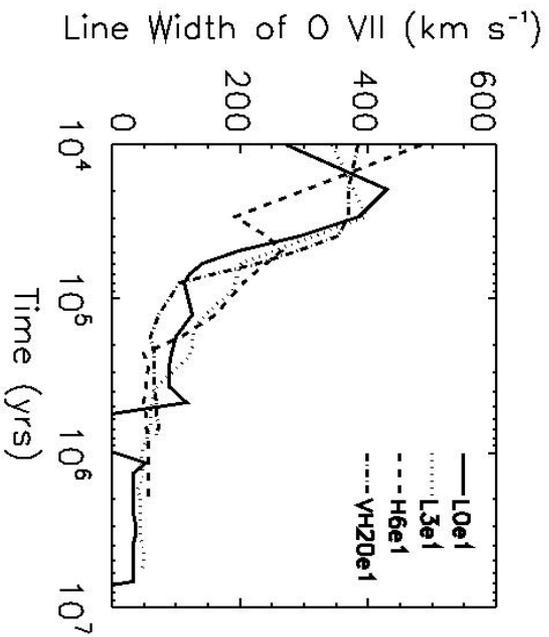

Fig. 6b

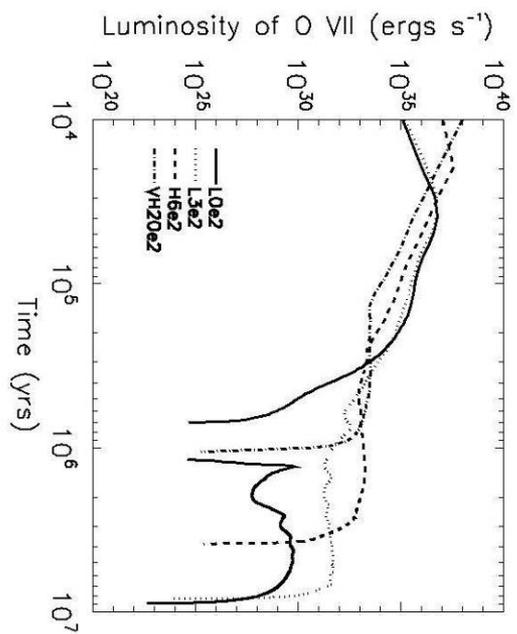

Fig. 6c

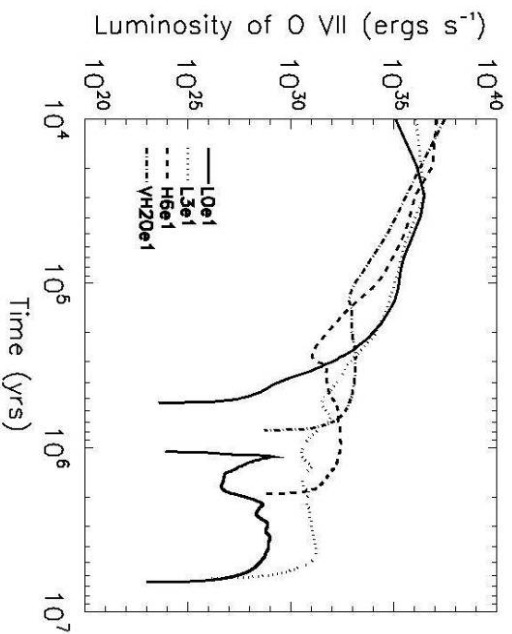

Fig. 6d

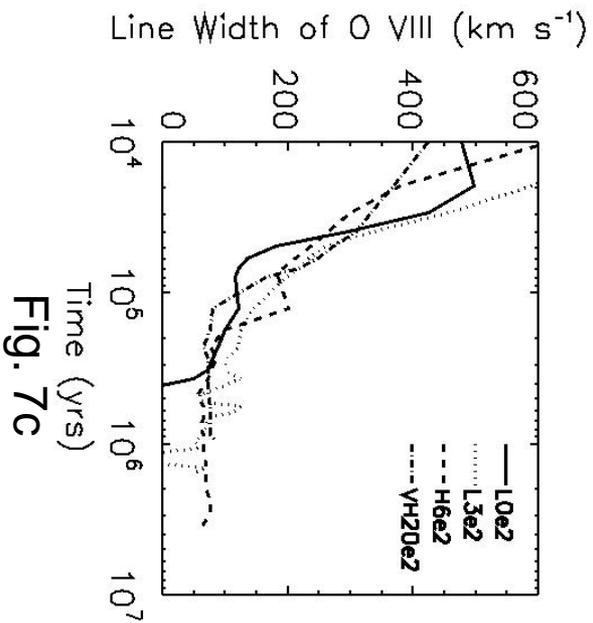

Fig. 7a

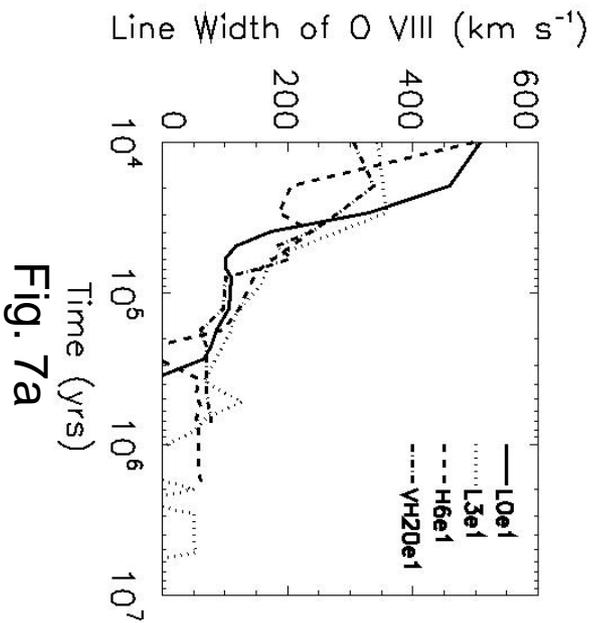

Fig. 7b

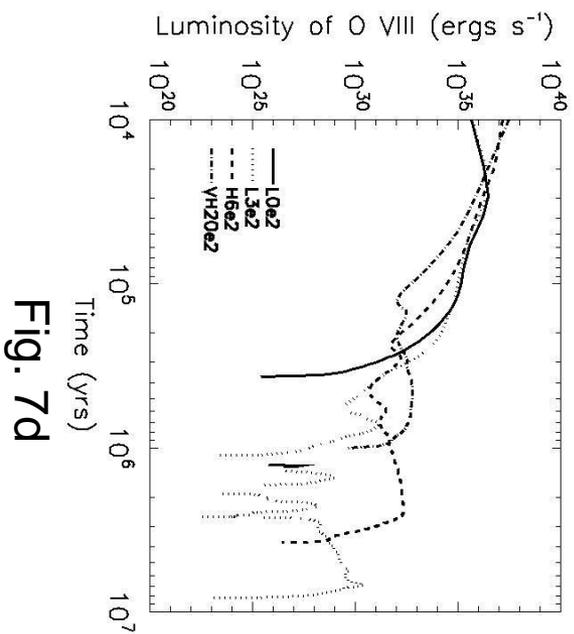

Fig. 7c

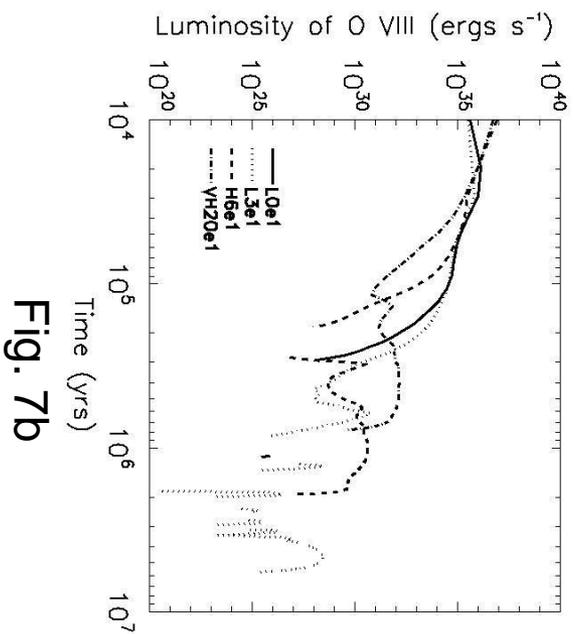

Fig. 7d

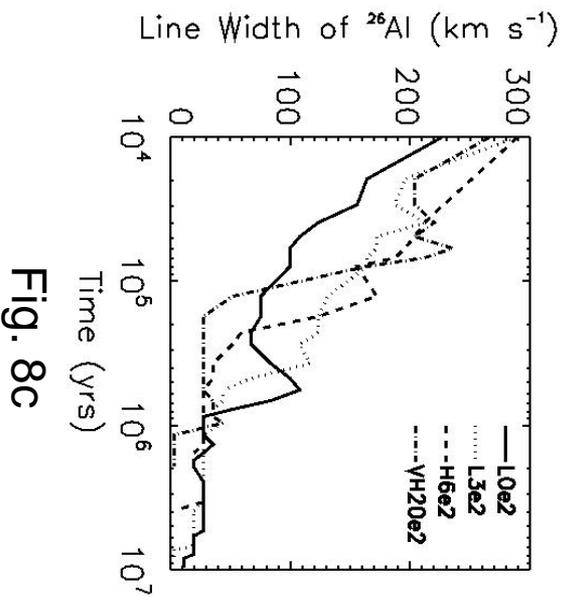

Fig. 8c

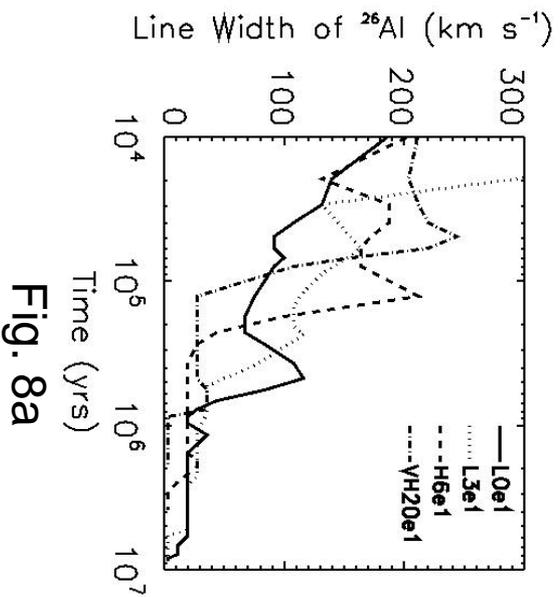

Fig. 8a

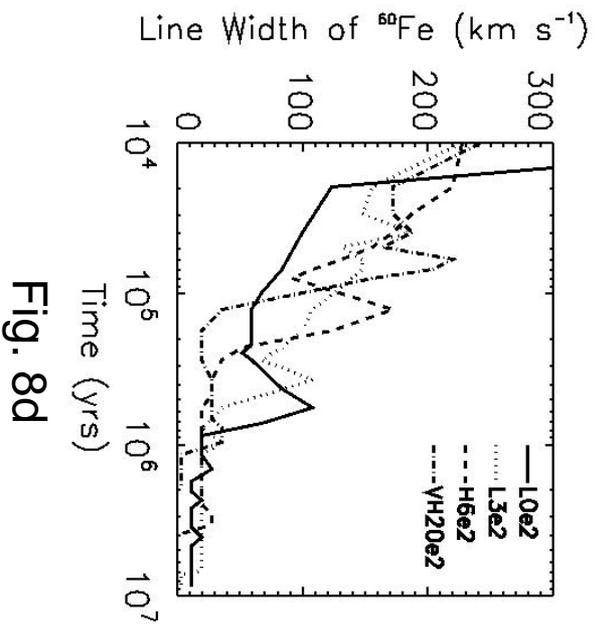

Fig. 8d

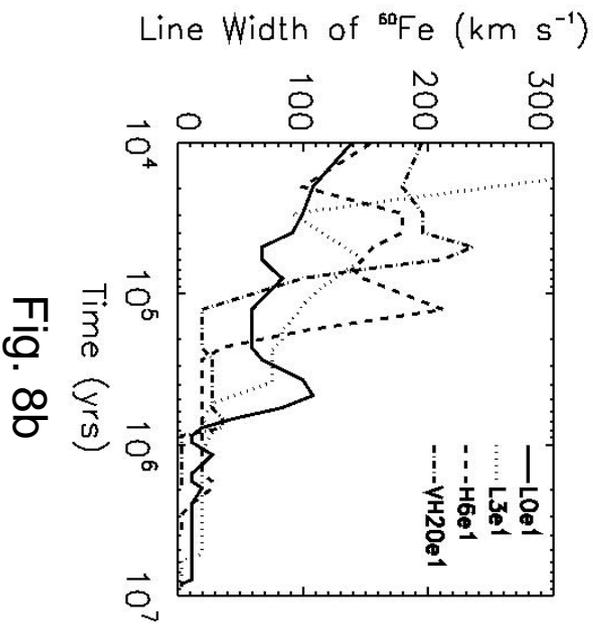

Fig. 8b